\documentclass[12pt]{iopart}

\usepackage{iopams}
\usepackage{setstack}
 
\usepackage[dvips]{graphicx}

\begin{document}

\title[Strange Resonance Production in p+p
and Au+Au Collisions at RHIC Energies]{Strange Resonance
Production in p+p and Au+Au Collisions at RHIC Energies}

\author{Christina Markert \dag\ for the STAR collaboration
\footnote[2]{For the full author list and acknowledgements see
Appendix "Collaborations" in this volume.}}

\address{\dag\ Physics Department, Yale University, New Haven, CT 06520,
USA}

\begin{abstract}
Resonance yields and spectra from elementary p+p and Au+Au
collisions at $\sqrt{s_{\rm NN}} = $ 200 GeV from the STAR
experiment at RHIC are presented and discussed in terms of
chemical and thermal freeze-out conditions. Thermal models do not
adequately describe the yields of the resonance production in
central Au+Au collisions. The approach to include elastic hadronic
interactions between chemical freeze-out and thermal freeze-out
suggests a time of $\Delta \tau>$5~fm/c.
\end{abstract}

\vspace{-0.5cm}

\section{Introduction}
Short-lived resonances and their decay products may interact
strongly with the hadrons from the fireball medium from the time
of chemical freeze-out until the system breaks up at kinetic
freeze-out. In order to try to understand the evolution and
expansion of this hot and dense medium we compare particle
resonance yields and spectra from elementary p+p and heavy ion
collisions. The observed decrease of the strange resonance to
non-resonance particle ratio in Au+Au collisions compared to p+p
collisions suggests that medium effects are indeed present in
heavy ion collisions. These ratios, along with a knowledge of the
resonance lifetime and interaction cross-sections, allow us to
place limits on the chemical freeze-out temperature and the
duration of the hadronic interactions in these collisions.
%Further, by studying the ratios as a function of centrality in
%Au+Au collisions we can determine the effect of the collision
%volume on the lifetime interval.

\section{Data Analysis}

The K(892), $\Sigma$(1385) and $\Lambda$(1520) resonances are
reconstructed by measuring their decay daughters with the STAR
{\em Time Projection Chamber} (TPC). Charged decay particles are
identified via energy loss ({\em dE/dx}) and their measured
momenta (K(892)$\rightarrow$K + $\pi$ and
$\Lambda$(1520)$\rightarrow$p + K). The neutral strong decays such
as $\Lambda$s from a $\Sigma$(1385) decay are reconstructed via
topological analysis ($\Lambda$($\rightarrow$p+$\pi$) + $\pi$).
The resonance signal is obtained by the invariant mass
reconstruction of each daughter combination and subtraction of the
combinatorial background calculated by mixed event or like-sign
techniques (figure~\ref{sigmapp}) \cite{gau03}. The resonance
ratios, spectra and yields are measured at mid-rapidity. The
central trigger selection for Au+Au collisions takes the 5\% of
most central inelastic interactions. The setup for the
proton+proton interaction is a minimum bias trigger.

%\begin{table}[htb]
%\begin{center}
%\begin{tabular}{lllll}
%\hline
%\Particle & mass [MeV/c$^{2}$]& width [MeV/c$^{2}$]& lifetime [fm/c]& decay channel \\
 %\%&(MeV/ c$^{2}$) & (MeV/ c$^{2}$) &(fm/ c) & \\
%\\hline
%\K(892) & 896.1 $\pm$ 0.27 &  50.7 $\pm$ 0.6 & 3.89 & K + $\pi$ \\
%\%$\phi$(1020) & 1019.417 $\pm$ 0.014 &  4.458 $\pm$ 0.032 &  44.6 & $K^{+}$  + $K^{-}$ \\
%\$\Sigma$(1385)$^{+}$ & 1382.8 $\pm$ 0.4 &  35.8 $\pm$ 0.8 & 5.5 & $\Lambda$($\rightarrow$p+$\pi$) + $\pi^{+}$ \\
%\$\Sigma$(1385)$^{-}$ & 1387.2 $\pm$ 0.5 &  39.4 $\pm$ 2.1 & 5.0 & $\Lambda$($\rightarrow$p+$\pi$) + $\pi^{-}$\\
%\$\Lambda$(1520)   & 1519.5 $\pm$ 1.0 &  15.6 $\pm$ 1.0 & 12.6 & p + K \\
%\ \hline
%\% \vspace{-2cm}
%\\end{tabular}
%\\end{center}
%\\vspace{-0.4cm}
%\ \caption[]{Selected resonances from PDG \cite{pdg98}}
%\\label{tab1} \label{resotable}
%\end{table}

\section{Strange Resonances in p+p Collisions at $\sqrt{s_{\rm NN}} = $ 200 GeV}

\begin{figure}[htb]
\centering
\includegraphics[width=0.45\textwidth]{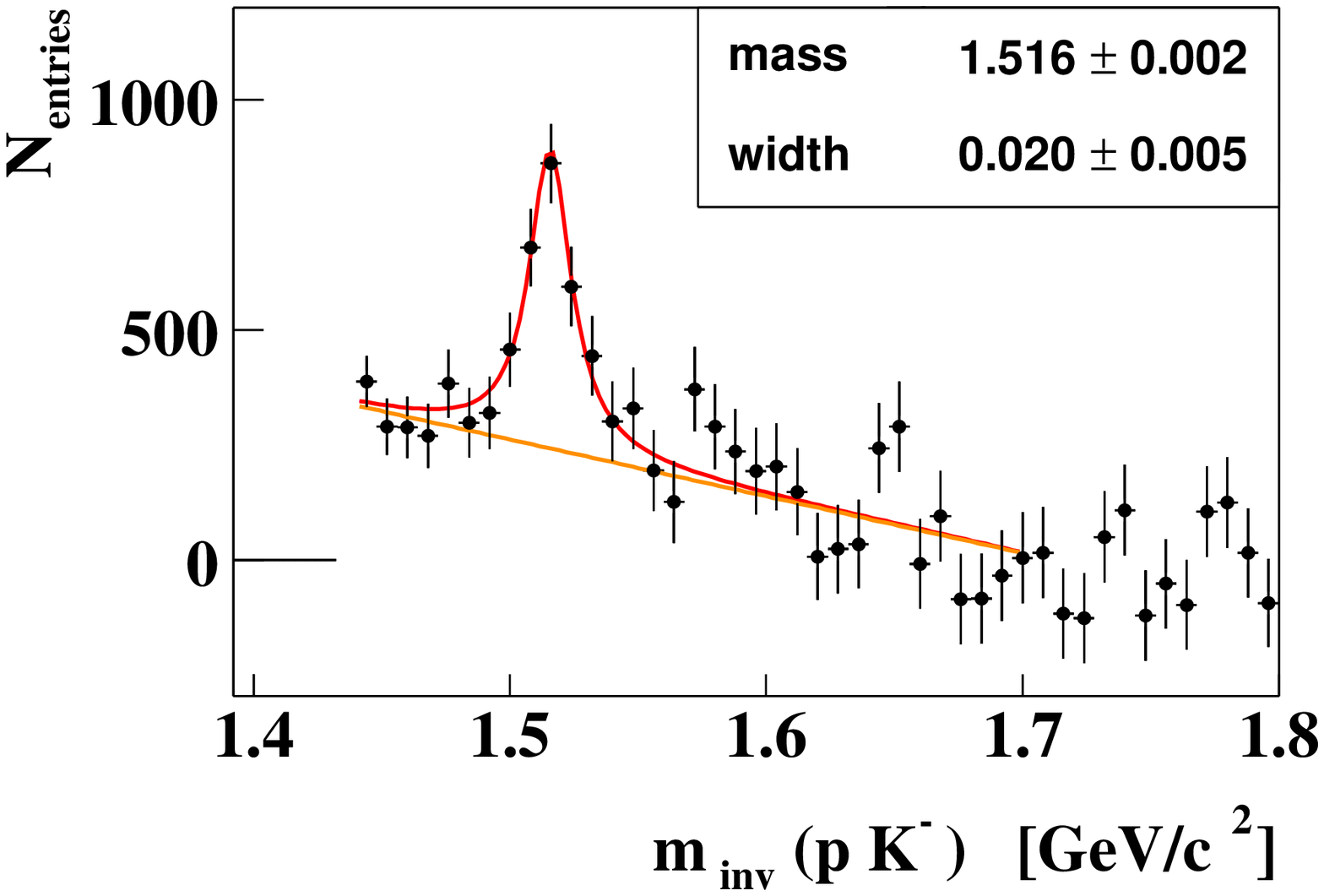}
\includegraphics[width=0.48\textwidth]{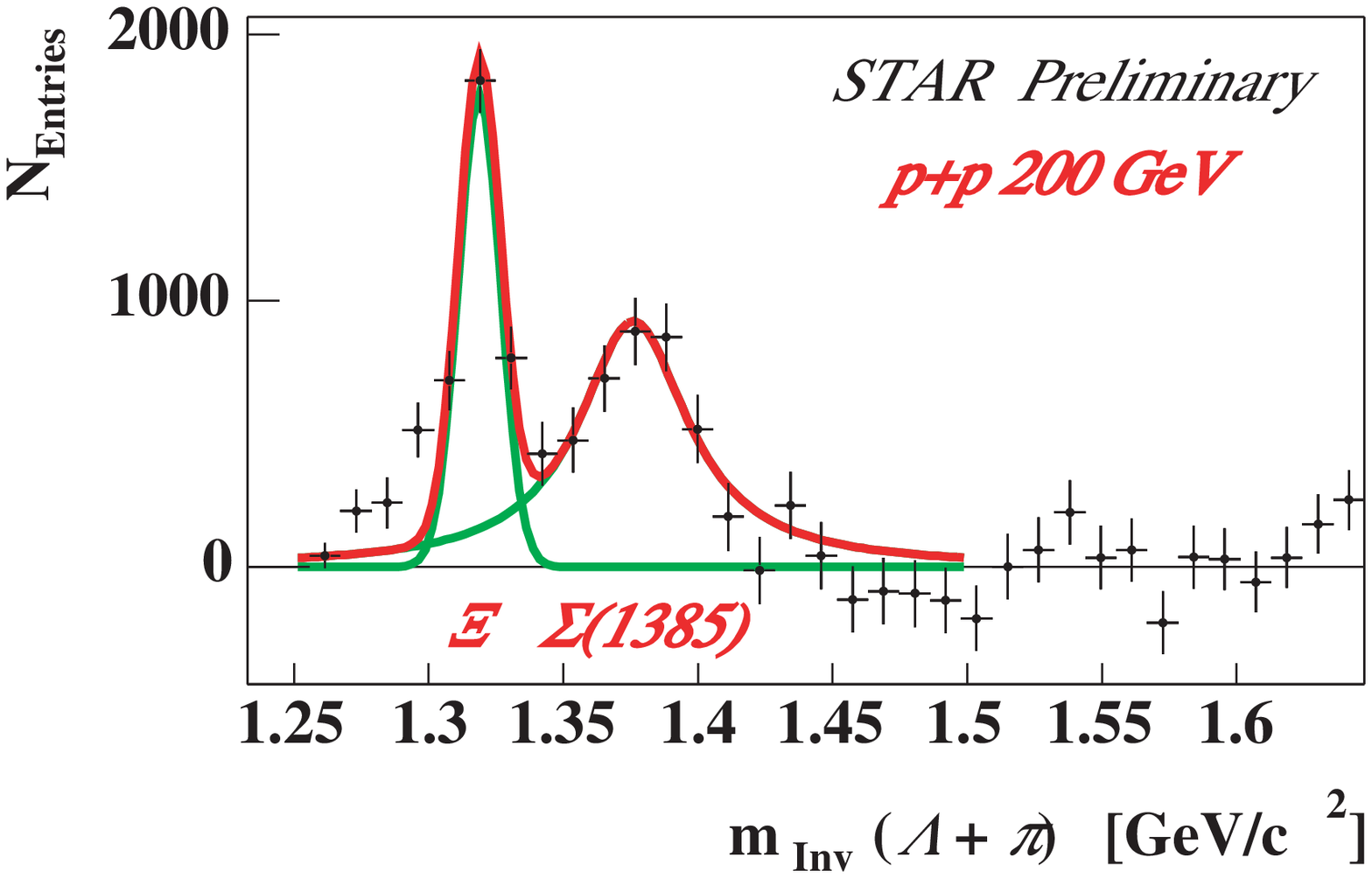}
\vspace{-0.4cm}
 \caption{Invariant mass distributions in p+p
collisions at $\sqrt{s_{\rm NN}}=$200 GeV after mixed-event
background subtraction for $\Lambda$(1520) $\rightarrow$ p+$K^{-}$
(left panel) and $\Sigma$(1385) $\rightarrow$ $\Lambda$+$\pi$
(right panel). The $\Xi$ and the $\Sigma$(1385) share the same
decay channel. Hence we also observe a strong $\Xi$ peak in the
$\Lambda$+$\pi$ decay channel.}
 \label{sigmapp}
\vspace{-0.1cm}
\end{figure}

The integrated invariant mass spectrum at mid-rapidity in p+p
collisions after background subtraction are shown in
figure~\ref{sigmapp} for $\Lambda$(1520) and
$\Sigma$(1385)$^{+/-}$ \cite{sal03}. The mass and width are in
agreement with the values from the PDG \cite{pdg98}, once the
expected contributions of the momentum resolution and detector
acceptance are included.

\vspace{-0.4cm}

\begin{figure}[htb]
\begin{minipage}[b]{0.33\linewidth}
 \centering
\includegraphics[width=1.0\textwidth]{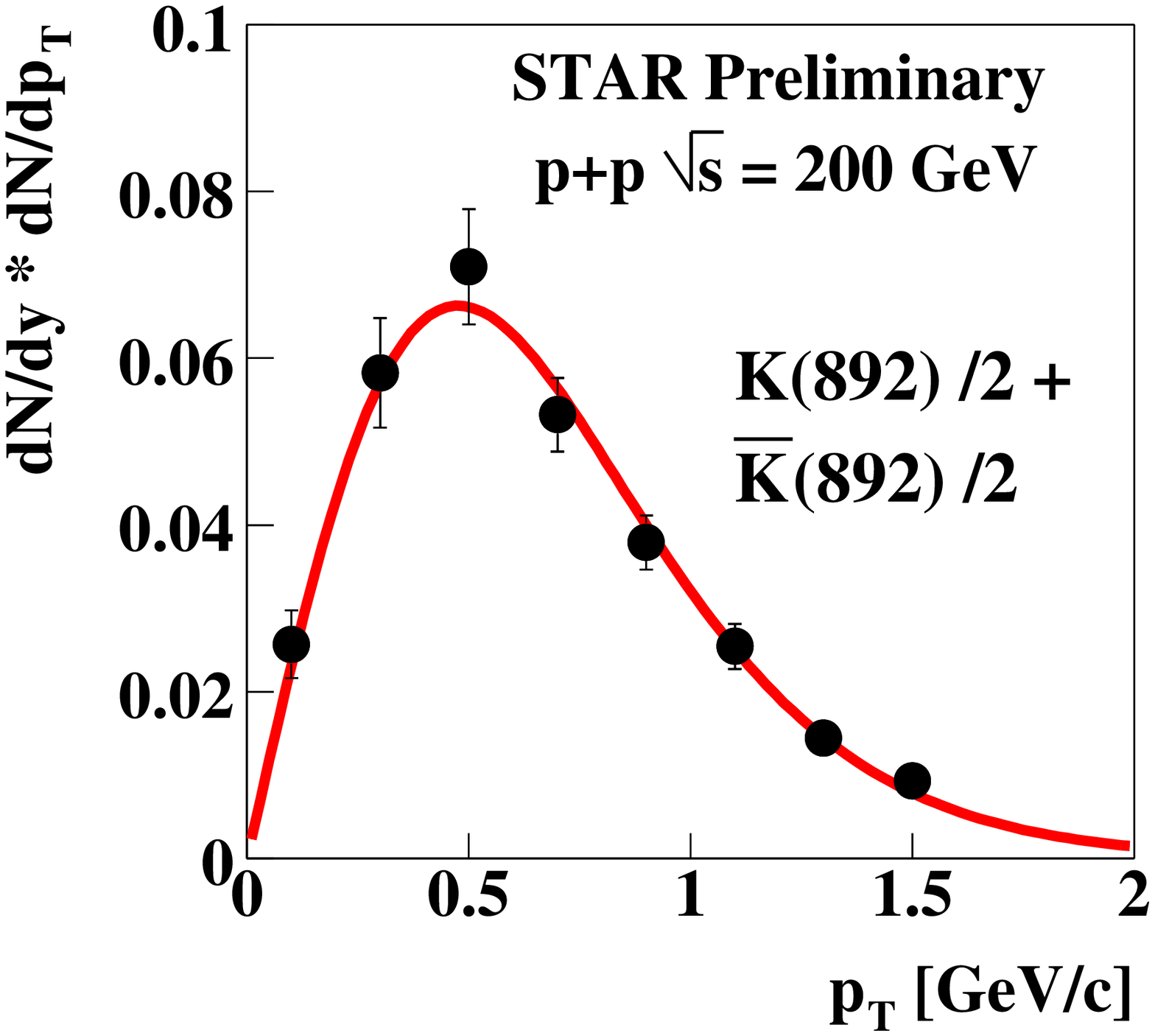}
 %\caption{This is a caption.}
%\vspace{1cm}
% \label{lampion}
 \end{minipage}
% \hspace{-0.5cm}
 \begin{minipage}[b]{0.33\linewidth}
 \centering
 \includegraphics[width=1.0\textwidth]{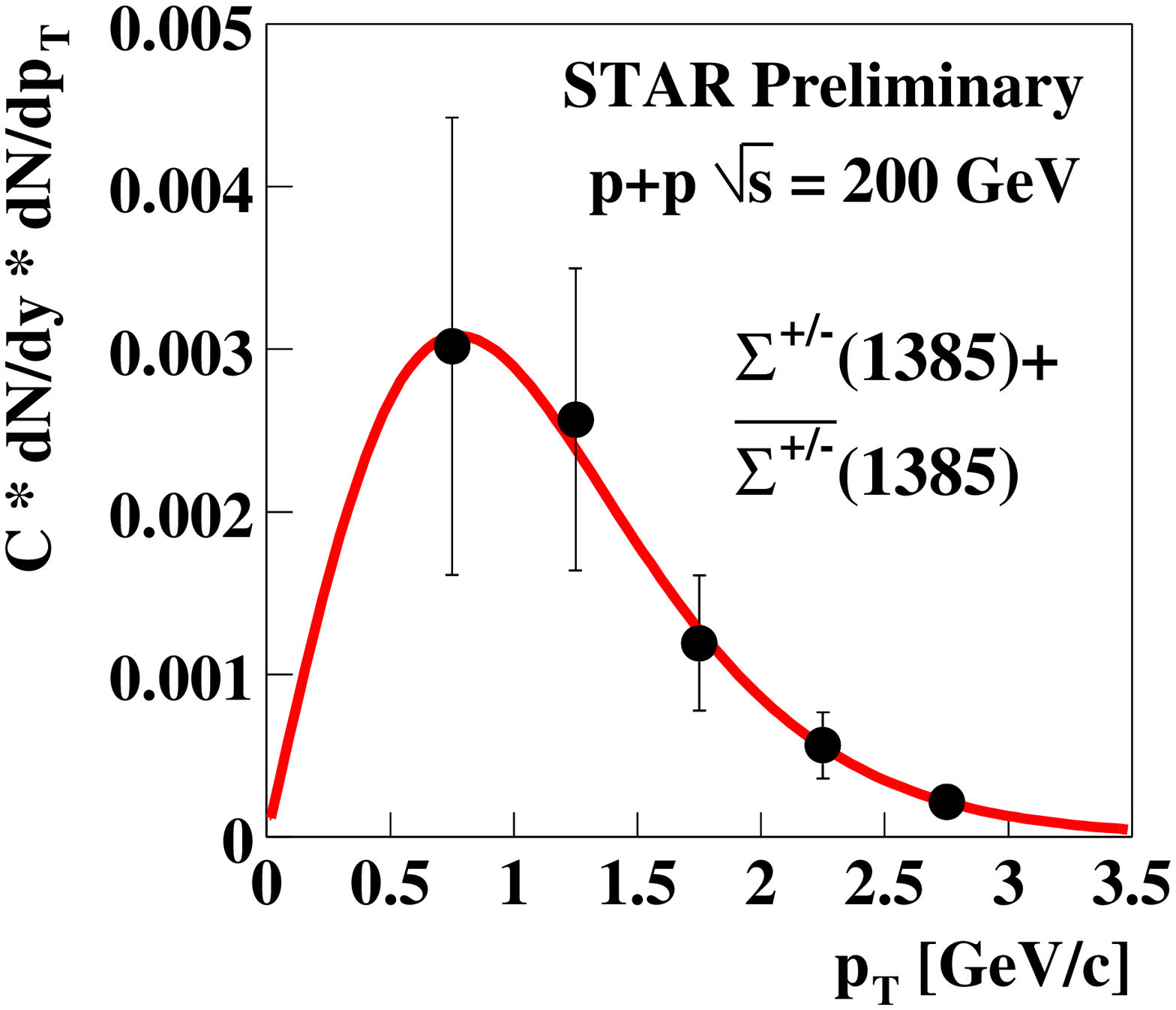}
 %\caption{This is a caption.}
% \label{lamlam}
\end{minipage}
 \begin{minipage}[b]{0.33\linewidth}
 \centering
 \includegraphics[width=1.0\textwidth]{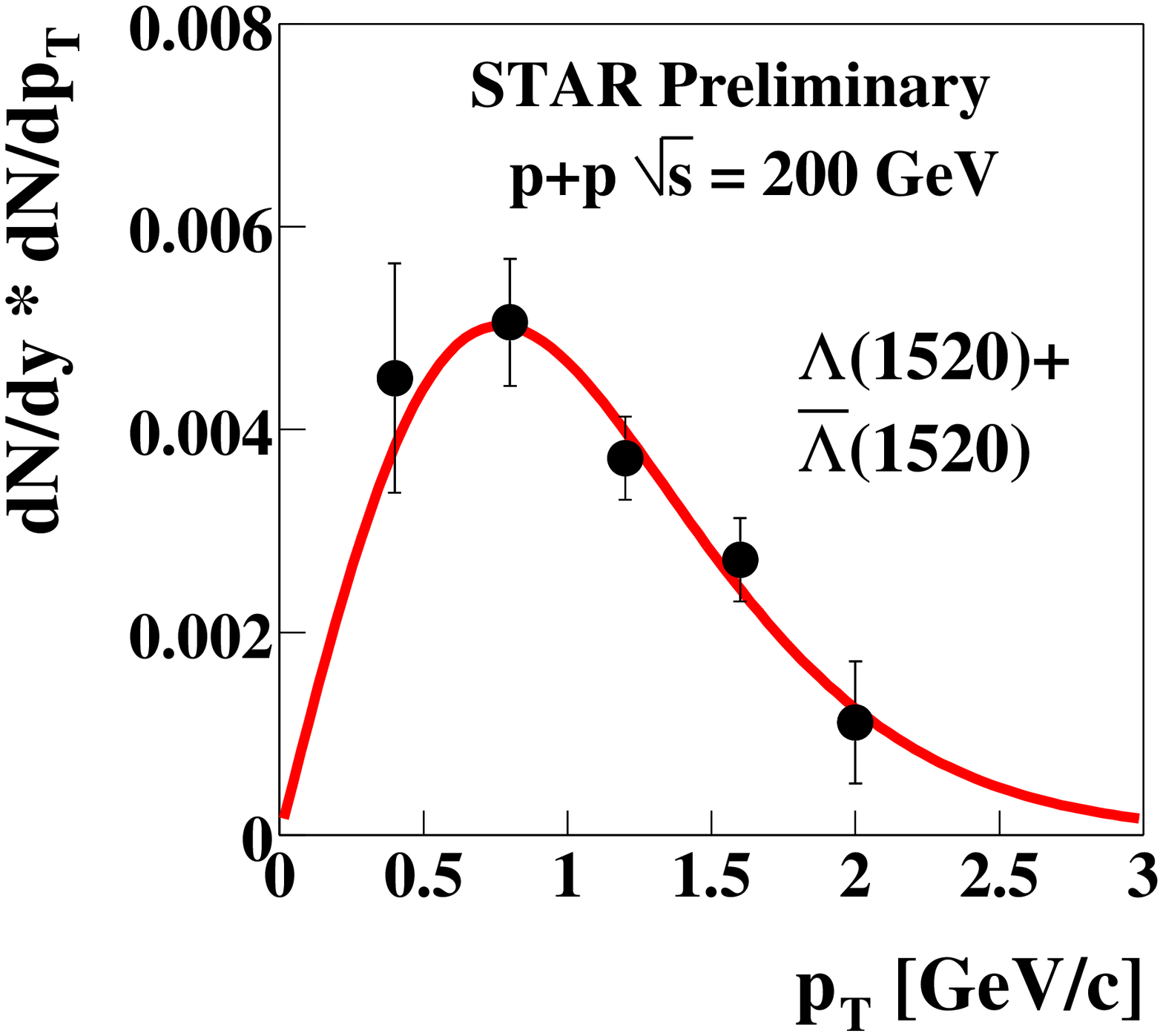}
 %\caption{This is a caption.}
\end{minipage}
 \vspace{-0.8cm}
 \caption{p$_{\rm T}$ distributions for K(892) \cite{zha03},
$\Sigma$(1385) and $\Lambda$(1520) for p+p collisions. The errors
for K(892) and $\Lambda$(1520) are statistical error only for the
$\Sigma$(1385) the systematic errors are included. The systematic
errors are 10-15\%.}
 \label{ptpp}
 \vspace{-0.4cm}
\end{figure}

\begin{figure}[htb]
\begin{minipage}[b]{0.5\linewidth}
 \centering
\includegraphics[width=1.0\textwidth]{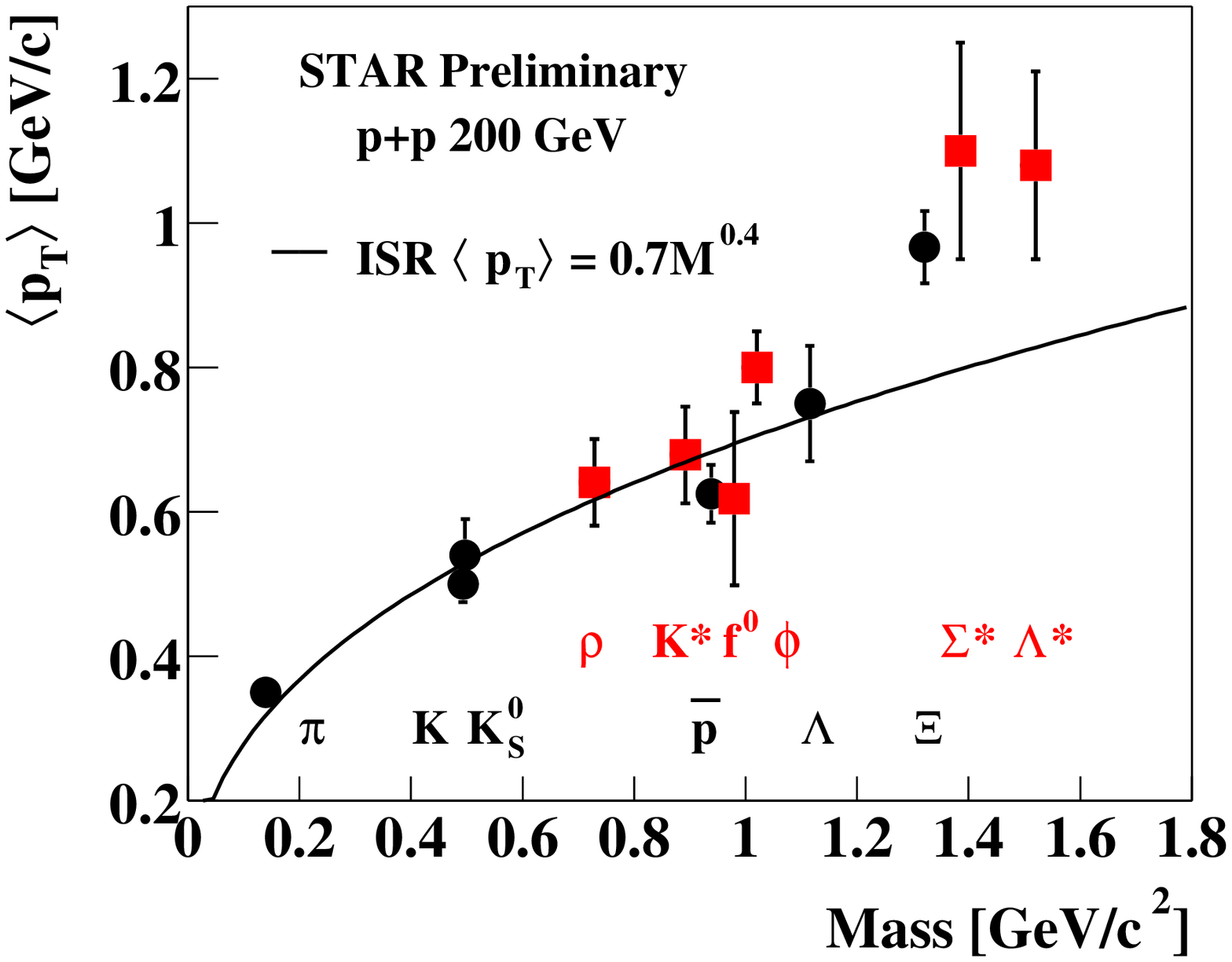}
 \end{minipage}
 \begin{minipage}[b]{0.5\linewidth}
 \centering
 \includegraphics[width=0.9\textwidth]{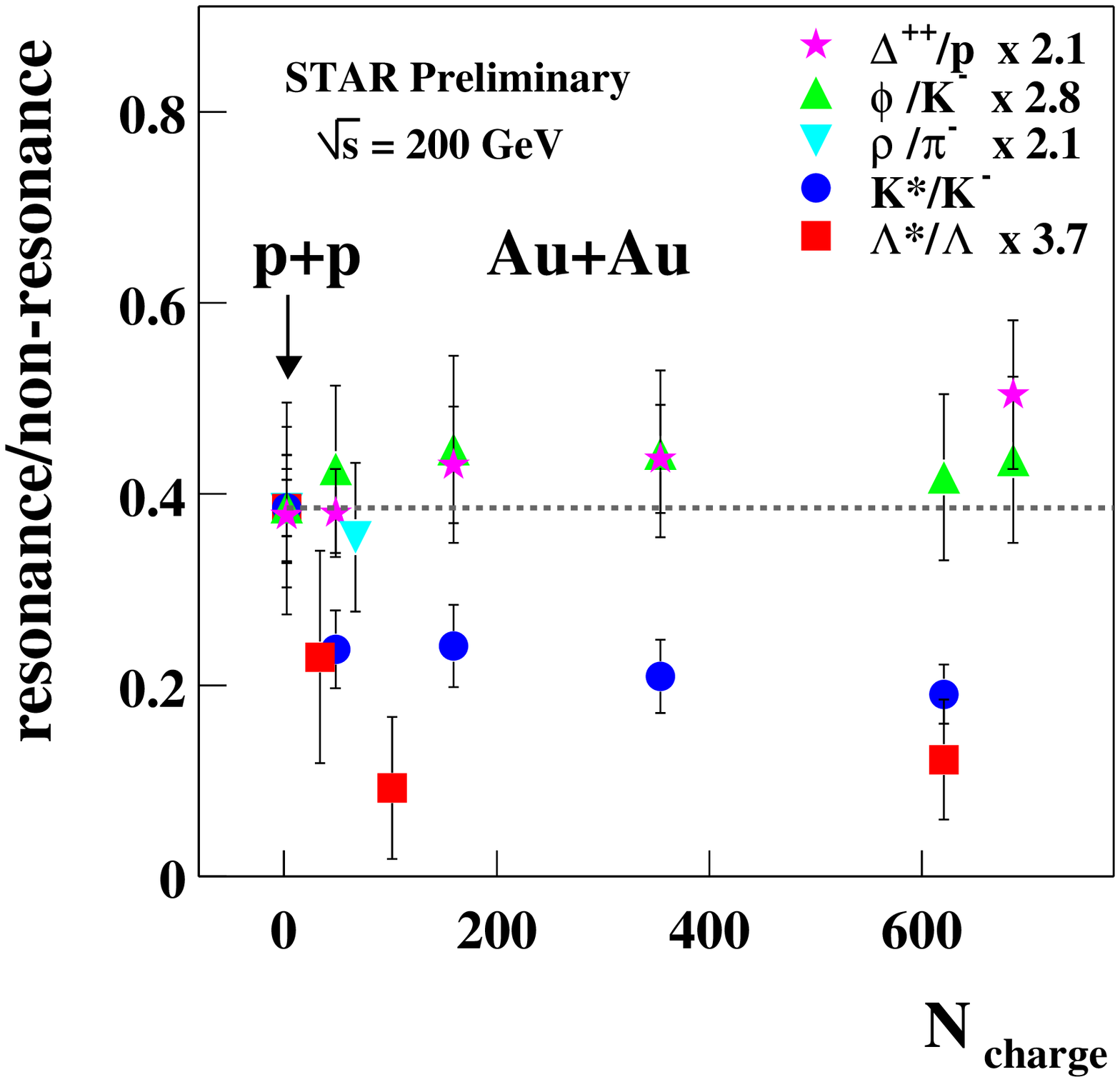}
 \vspace{-0.3cm}
\end{minipage}
\caption{Left: $\langle$p$_{\rm T}$$\rangle$ as a function of
particle mass in p+p collisions at $\sqrt{s_{\rm NN}}=$200 GeV.
The black curve is a fit to
 measured ISR p+p data ($\pi$, K and p) a collision energy of $\sqrt{s_{\rm NN}}=$26 GeV \cite{bou76}. Right: Resonance/non-resonance ratios of
$\phi$/K$^{-}$ \cite{ma03}, $\Delta^{++}$/p \cite{zha04},
$\rho/\pi$ \cite{fac03}, K(892)/K$^{-}$ \cite{zha03} and
$\Lambda$(1520)/$\Lambda$ \cite{gau03,mar03} for p+p and Au+Au
collisions at $\sqrt{s_{\rm NN}} = $ 200 GeV. The ratios are
normalized to the K*(892)/K$^{-}$ p+p ratio. Statistical and
systematic errors are included.}
 \label{part}
 %\vspace{-0.2cm}
\end{figure}

The transverse momentum distributions of K(892) \cite{zha03},
$\Sigma$(1385) \cite{sal03} and $\Lambda$(1520) from p+p
collisions are shown in figure~\ref{ptpp}. The inverse slope
parameter, T, and the mean transverse momentum ,$p_{\rm T}$, are
obtained by an exponential fit (see table below). The transverse
momentum coverage for the resonance yields is approximately
80-95\%. The mean transverse momentum as a function of mass in
figure~\ref{part} (left panel) shows a mass dependence for stable
(circles) and resonance (squares) particles. A fit to the ISR data
shown by the curve includes $\pi$, K and p from p+p collisions at
$\sqrt{s_{\rm NN}}=$26 GeV \cite{bou76} and predicts a much less
pronounced mass dependence in good agreement with the $\pi$, K and
p data from $\sqrt{s_{\rm NN}}=$200 GeV. The resonances with mass
higher than 1 GeV/c$^{2}$ and the $\Xi$ indicate a stronger
$\langle$p$_{\rm T}$$\rangle$ dependence which is not represented
by the fit to the ISR data.

\begin{table}[htb]
%\begin{center}
%\vspace{-3cm}
\hspace{1cm}
\begin{minipage}[h]{0.6\linewidth}

\begin{tabular}{lll}
\hline
Particle & T [MeV] &  $\langle p_{\rm T} \rangle$ [GeV] \\
\hline
K(892)$^{0}$ & 223 $\pm$ 9 & 0.68 $\pm$ 0.03 $\pm$ 0.03 \\
$\Sigma$(1385) & 380 $\pm$ 50  & 1.10$\pm$ 0.02 $\pm$ 0.01 \\
 $\Lambda$(1520) & 345 $\pm$ 42 & 1.08 $\pm$ 0.09$\pm$ 0.11  \\
 \hline
\end{tabular}
\end{minipage}

%\end{center}
\vspace{-2.5cm} \hspace{8cm}
 \begin{minipage}[h]{0.4\linewidth}
%\vspace{-3cm}
%\hspace{4cm}
 \caption{T and $\langle p_{\rm T} \rangle$ for resonances in p+p interactions.}
 \end{minipage}
 \label{resotable}
\end{table}
\vspace{0.7cm}
%\vspace{3cm}

%\vspace{-1cm}

\section{Strange Resonances in Au+Au Collisions at $\sqrt{s_{\rm NN}} = $ 200 GeV}

For a comparison of resonance production in different collision
systems, the resonances are divided by non-resonance particle
yields to eliminate the need for a volume and energy
normalization. Figure~\ref{part} right shows the different
resonance to stable particle ratios for p+p and Au+Au collisions
as a function of the mid-rapidity charged particle yields N$_{\rm
charge}$. All the ratios are normalized such that their p+p ratio
equals that of the K(892)/K$^{-}$ p+p value. The K(892)/K$^{-}$
and $\Lambda$(1520)/$\Lambda$ ratios decreases from p+p to Au+Au
collision systems. This behavior shows that the resonance
production in Au+Au is not a simple superposition of p+p
interactions. It is also an indication that the surrounding
extended medium of a Au+Au collision has an influence on the
resonances and/or their decay particles. The thermal model is able
to fit the ratios of the stable particles \cite{pbm01}. The
results from these fits predicts $\Lambda$(1520)/$\Lambda$ = 0.07
and K(892)/K$^{-}$=0.32 at a chemical freeze-out temperature of
T=170 MeV . The measured value for central Au+Au collisions at
mid-rapidity are $\Lambda$(1520)/$\Lambda$ = 0.034 $\pm$ 0.011
$\pm$ 0.013 and K(892)/K$^{-}$ = 0.19 $\pm$ 0.05, significantly
lower than the thermal model predictions. A thermal description of
these measured ratios would lead to a 30-50 MeV smaller chemical
freeze-out temperature.

\section{Time Scale for Au+Au collisions}

\vspace{-0.3cm}

In-medium effects such as elastic interactions of the decay
particles with other particles in the created medium (mostly
pions) may result in a signal loss. Regeneration will have the
reverse effect. Microscopic model calculations (UrQMD
\cite{ble02,ble03}) predict $\Lambda$(1520)/$\Lambda$ = 0.03 and
K(892)/K$^{-}$ = 0.25 (lifetime: $\tau_{K(892)}$ = 4 fm/c and
$\tau_{\Lambda(1520)}$ = 13 fm/c). Meanwhile the UrQMD calculation
for $\Delta^{++}$/p = 0.28 is compatible with the data,
$\Delta^{++}$/p = 0.240 $\pm$ 0.037. Since the $\Delta^{++}$/p
ratio in Au+Au is nearly the same as it is for p+p collisions,
either the regeneration of the signal is of the same order as the
rescattering or there is no rescattering. The $\Delta^{++}$ has a
short lifetime of 1.3 fm/c therefor the rescattering of the decay
daughters is expected to be large. Due to the long life-time of
the $\phi$ resonance (44 fm/c) we would expect no significant
signal loss by rescattering of the daughters since most of the
decays happen outside of the fireball, which is in agreement with
our observation. Using the measured values of K(892)/K and
$\Lambda$(1520)/$\Lambda$ a thermal model, including rescattering,
but no regeneration, predicts the lifetime interval between
chemical and kinetic freeze-out to be $>$5 fm/c.
\cite{tor01,mar02}. The decrease of the $\Lambda$(1520)/$\Lambda$
and K(892)/K$^{-}$ ratio from p+p to Au+Au collisions occurs
already for very peripheral collisions (50-80\%) and remains
nearly constant up to the most central collisions (5\%). In terms
of the lifetime of the system between chemical and thermal
freeze-out, $\Delta\tau$, these result suggests the same
$\Delta\tau$ for peripheral and central Au+Au collisions.

\vspace{-0.3cm}

\section{Conclusions}

\vspace{-0.3cm}

Resonances are a unique tool to probe the time span $\Delta\tau$
between chemical and thermal freeze-out in heavy ion collisions.
The measured $\Lambda$(1520)/$\Lambda$ and K(892)/K$^{-}$ together
with a thermal model and rescattering suggest a lower limit of
$\Delta\tau$ $>$5 fm/c. Furthermore, the high mass particles show
a stronger mass dependence of $\langle$p$_{\rm T}$$\rangle$ in p+p
collisions than predicted from the stable particles below 1 GeV.

%Figure~\ref{mtpp} shows the transverse mass distribution of K(892)
%and  $\Lambda$(1520) where the inverse slope parameter T is
%obtained with $T_{K(892)}$ = 223 $\pm$ 9 MeV for p+p interactions
%and $T_{K(892)}$ = 350 $\pm$ 23 MeV for the 70-80\% most
%peripheral Au+Au collisions. This strong increase of 100 MeV in
%the inverse slope parameter is not obsereved for the stable
%particles.

%\subsection{Acknowledgments}
%I also would like to thank the STAR collaboration for support in
%presenting this data. And specially I would like to thank Sevil
%Salur, Ludovic Gaudichet, Patricia Fachini, Jingguo Ma, An Tai and
%Haibin Zhang.

\end{document}